\documentclass[twocolumn]{aastex631}

\usepackage{enumitem,amssymb}
\usepackage[utf8]{inputenc}
\usepackage{color}
\usepackage{graphicx}	
\usepackage{amsmath}	
\usepackage{amssymb}	
\usepackage{wasysym}
\usepackage{mathtools}
\usepackage{newtxtext,newtxmath}
\usepackage{appendix}

\newcommand{\hst}{\textit{HST}}
\newcommand{\jwst}{\textit{JWST}}
\newcommand{\lya}{Ly$\alpha$}

\newcommand{\flam}{$f_\lambda$}
\newcommand{\fnu}{$f_\nu$}

\received{November 18, 2022}

\submitjournal{ApJ}

\shorttitle{Spectral Templates for High-z Galaxies}
\shortauthors{Larson et al.}

\begin{document}

\title{Spectral Templates Optimal for Selecting Galaxies at $z > 8$ with \jwst}

\correspondingauthor{Rebecca L. Larson}
\email{rlarson@astro.as.utexas.edu}

\author[0000-0003-2366-8858]{Rebecca L. Larson}
\altaffiliation{NSF Graduate Fellow}
\affiliation{The University of Texas at Austin, Department of Astronomy, Austin, TX, United States}

\author[0000-0001-6251-4988]{Taylor A. Hutchison}
\altaffiliation{NSF Graduate Fellow}
\altaffiliation{NASA Postdoctoral Fellow}
\affiliation{Department of Physics and Astronomy, Texas A\&M University, College Station, TX, 77843-4242 USA}
\affiliation{George P. and Cynthia Woods Mitchell Institute for Fundamental Physics and Astronomy,\\ Texas A\&M University, College Station, TX, 77843-4242 USA}
\affiliation{Astrophysics Science Division, NASA Goddard Space Flight Center, 8800 Greenbelt Rd, Greenbelt, MD 20771, USA}

\author[0000-0002-9921-9218]{Micaela Bagley}
\affil{The University of Texas at Austin, Department of Astronomy, Austin, TX, United States}

\author[0000-0001-8519-1130]{Steven L. Finkelstein}
\affil{The University of Texas at Austin, Department of Astronomy, Austin, TX, United States}

\author[0000-0003-3466-035X]{L. Y. Aaron\ Yung}
\altaffiliation{NASA Postdoctoral Fellow}
\affiliation{Astrophysics Science Division, NASA Goddard Space Flight Center, 8800 Greenbelt Rd, Greenbelt, MD 20771, USA}

\author[0000-0002-6748-6821]{Rachel S. Somerville}
\affiliation{Center for Computational Astrophysics, Flatiron Institute, 162 5th Avenue, New York, NY, 10010, USA}

\author[0000-0002-3301-3321]{Michaela Hirschmann}
\affiliation{Institute of Physics, Laboratory of Galaxy Evolution, Ecole Polytechnique Fédérale de Lausanne (EPFL), Observatoire de Sauverny, 1290 Versoix, Switzerland}

\author[0000-0003-2680-005X]{Gabriel Brammer}
\affil{Cosmic Dawn Center (DAWN), Niels Bohr Institute, University
of Copenhagen, Jagtvej 128, København N, DK-2200, Denmark}

\author[0000-0002-4884-6756]{Benne W. Holwerda}
\affil{Physics \& Astronomy Department, University of Louisville, 40292 KY, Louisville, USA}

\author[0000-0001-7503-8482]{Casey Papovich}
\affiliation{Department of Physics and Astronomy, Texas A\&M University, College Station, TX, 77843-4242 USA}
\affiliation{George P. and Cynthia Woods Mitchell Institute for Fundamental Physics and Astronomy,\\ Texas A\&M University, College Station, TX, 77843-4242 USA}

\author[0000-0003-4965-0402]{Alexa M.\ Morales}
\affiliation{The University of Texas at Austin, Department of Astronomy, Austin, TX, United States}

\author[0000-0003-3903-6935]{Stephen M.~Wilkins} %
\affiliation{Astronomy Centre, University of Sussex, Falmer, Brighton BN1 9QH, UK}
\affiliation{Institute of Space Sciences and Astronomy, University of Malta, Msida MSD 2080, Malta}

\begin{abstract}

The selection of high-redshift galaxies often involves spectral energy distribution (SED) fitting to photometric data, an expectation for contamination levels, and measurement of sample completeness -- all vetted through comparison to spectroscopic redshift measurements of a sub-sample. The first {\it JWST} data is now being taken over several extragalactic fields, to different depths and across various areas, which will be ideal for the discovery and classification of galaxies out to distances previously uncharted. As spectroscopic redshift measurements for sources in this epoch will not be initially available to compare with the first photometric measurements of $z > 8$ galaxies, robust photometric redshifts are of the utmost importance. Galaxies at $z > 8$ are expected to have bluer rest-frame ultraviolet (UV) colors than typically-used model SED templates, which could lead to catastrophic photometric redshift failures. We use a combination of BPASS and {\sc Cloudy} models to create a supporting set of templates that match the predicted rest-UV colors of $z >$ 8 simulated galaxies.  We test these new templates by fitting simulated galaxies in a mock catalog \citet{yung22}, which mimic expected field depths and areas of the {\it JWST} Cosmic Evolution Early Release Science Survey (CEERS: m$_{5\sigma} \sim$ 28.6 over $\sim$100 arcmin$^2$; \citealt{finkelstein22b,bagley22}).  We use EAZY to highlight the improvements in redshift recovery with the inclusion of our new template set and suggest criteria for selecting galaxies at $8 < z < 10$ with {\it JWST}, providing an important test case for observers venturing into this new era of astronomy.

\end{abstract}

\section{Introduction} \label{sec:intro}

  As we enter the {\it James Webb Space Telescope} (\jwst) era of high-redshift galaxy studies, the early Universe has been opened up to discovery. Deep 1-5 $\mu$m \jwst\ imaging paired with {\it Hubble Space Telescope} (\hst) optical imaging allows galaxy selection via their Ly$\alpha$-breaks (below which intergalactic hydrogen absorbs the rest-frame ultraviolet [UV] light emitted from distant galaxies). The \jwst\ coverage allows detected galaxies to have both multiple “dropout” bands (non-detections blue-ward of the break) and multiple detection bands (significant detections red-ward of the break), substantially improving the discovery of galaxies in the reionization epoch ($z>7$).  The biggest advances with \jwst\ data lie at $z >$ 9, where \hst\ efforts could only see such galaxies in at 1-2 filters at $z \sim$ 9–10, and not at all at $z >$ 11.  Unsurprisingly, within days of the data being released, several studies identified tens of candidate galaxies at $z >$ 10 \citep[e.g.][]{adams22b, atek22b}, with a few at $z \gtrsim$ 12 \citep[][]{finkelstein22b, harikane22b,finkelstein22d}, and  even $z \sim$ 17 \citep[][]{donnan22}.  The observed number density of galaxy candidates is exceeding predictions \citep[][]{finkelstein22d}, with a variety of theoretical explanations already popping up exploring possible explanations ranging from dust-free stellar populations \citep[][]{ferrara22b} to extremely efficient star-formation \citep[][]{mason22, mirocha22}. 
 
The selection of high-redshift galaxies often involves spectral energy distribution (SED) fitting to photometric data, vetted through comparison to spectroscopic redshift measurements. The first \jwst\ data is now being taken over a variety of fields, to different depths and across various areas, which will be ideal for the discovery and classification of galaxies out to distances previously unobtainable. As statistically-significant spectroscopic redshift measurements for sources in this epoch will not be initially available to compare with the first photometric measurements of $z > 8$ galaxies, robust photometric redshifts are of the utmost importance. While photometric-redshift calculations at these high redshifts primarily measure the Lyman break, similar to color–color selection \citep[e.g.][]{steidel93, giavalisco04b, bouwens15b, bridge19}, SED fitting has the advantage that it simultaneously uses all available photometric information. This simplifies the selection process and results in a more inclusive sample of high-redshift candidates than color-color selection alone as it includes objects that might fall just outside color selection windows \citep[e.g.][]{mclure09, finkelstein10, finkelstein15c, bowler12, atek15, livermore17, bouwens19}. 
 
 To perform photometric redshift estimations many use  photometric-redshift (photo-z) codes such as EAZY\footnote{\href{https://github.com/gbrammer/eazy-photoz}{github.com/gbrammer/eazy-photoz}} \citep{brammer08}. These codes use all available photometry and compare to a series of SED templates, allowing nonlinear combinations of any number of provided templates. While these templates have been optimized to best match well-studied spectroscopic redshifts, the bulk of these spectroscopic measurements are at $z <$ 4. Thus the appropriateness of these templates for galaxies at higher redshifts, such as those of particular interest to \jwst\ surveys, is less well known.
 
 Galaxies at $z > 8$ are expected to have bluer rest-frame ultraviolet (UV) colors than typically-used model templates, which could lead to catastrophic photometric redshift failures. \citet{finkelstein22a} compared the native EAZY template set to their sample of $z =$ 6-8 galaxies \citep{finkelstein15}, and found that these templates did not span the full color range of their comparison sample (Figure 5 in their paper).  Specifically, while many of the included templates in EAZY were redder than these high-redshift galaxies, the bluest template was only as blue as their median high-redshift galaxy. It is expected that the $z>8$ galaxies that will be studied in depth with \jwst\ will have colors at least as blue as those $z=6-8$ galaxies discussed by \citet{finkelstein22a}. Due to expected young stellar populations at such early times in cosmic history, a decrease in metallicity at higher redshifts, and active star formation episodes these high-redshift galaxies likely have increasingly bluer colors. It is imperative that we use appropriate models in our SED fits to ensure the accuracy of our photometric redshifts. 
 
The selection of high-redshift galaxies is often made even more difficult due to the high rates of contamination from lower-redshift galaxies that mimic many of the same selection criteria. We must be looking into the best ways to reduce the contamination fraction in our candidate galaxy selection process. This is often done by utilizing a number of spectroscopic redshift measurements to calibrate photometric redshift accuracy, where we are able to get a measure of how disparate the actual vs recovered redshifts of galaxies are on average. Unfortunately, during the first years of \jwst\ data we will not have a significant number of spectroscopic redshifts available above $z\sim8$ with which to conduct this comparison. The time is now for building up samples of galaxies in the reionization era as upcoming galaxy legacy surveys, and ensuring accurate measurements of their redshifts. To enable these analyses, we use a catalog of simulated galaxies that are expected to be representative of those at $z>8$ and perform an SED-fitting process to determine the accuracy and coverage of current SED templates. We explore how the creation of a new suite of blue galaxy templates can improve our fits to these simulated galaxies, and discuss the best selection criteria for selecting high-redshift galaxies with \jwst. 

In \S 2 we discuss the simulated galaxy catalog which provides a robust sample of $z=8-10$ galaxies with which to test our SED-fitting templates and methods. In \S 3 we address the color-space that our simulated, and expected real high-z galaxies, occupy but which is not covered by existing galaxy templates in the EAZY software and the templates we created to span this gap. We then test the improvements to our photometric redshift fits from EAZY that these new templates enable in \S 4. We also explore the robustness of our photometric redshift fits to these simulated galaxies when placed at depths equal to those predicted from one of the \jwst\ Early Release Science surveys with public data access in the first months of observations: the Cosmic Evolution Early Release Science Survey (CEERS: m$\sim$28.6, 100 arcmin$^2$, PI Finkelstein, \citealt{bagley22}) in \S 5. We also provide some suggested criteria for selecting galaxies at $z>8$ with \jwst\ in \S 6, which minimizes the contamination from low-redshift interlopers while maintaining a successful recovery-rate of target high-z galaxies. We then present our conclusions in \S 7. For this paper we express all magnitudes in the AB system \citep[][]{oke83} unless otherwise noted. In this paper we assume the latest Planck flat $\Lambda$CDM cosmology with $H_0$ = 67.36 km s$^{-1}$ Mpc$^{-1}$, $\Omega_m$ = 0.3153, and $\Omega_{\Lambda}$ = 0.6847 \citep[][]{planck20}.

\section{Simulating the High-Redshift Universe}

As the first {\it JWST} data are released and attention turns to the $z >$ 8 Universe, we explore whether previously-used templates are blue enough to match the expected colors of $z >$ 8 galaxies. We note that while {\it HST} did discover some galaxies at $z =$ 9--10, most were fairly massive \citep[e.g.][]{tacchella22} and thus might not have colors indicative of the bulk of the lower-mass population which {\it JWST} will study. As data from \jwst\ is only just starting to be taken, we explore the expected color-space of these galaxies by using a simulated catalog. 

In this work, we adopt three realizations of a modified version of simulated lightcones with footprints overlapping the observed EGS field as presented in \citet{yung22} and \citet{somerville21}. Each of these lightcones spans 782 arcmin$^2$ with dimensions of $17$ arcmin $\times$ 46 arcmin, containing galaxies in $0\lesssim z\lesssim 10$ and resolving galaxies down to $M_* \sim 10^7 {\rm M}_\odot$. The mock lightcone is constructed based on dark matter halos extracted from the Bolshoi-Planck $n$-body cosmological simulation \citep{klypin16} using the \texttt{lightcone} package provided as part of the the publicly-available \textsc{UniversaMachine} code \citep{behroozi19, behroozi20}. These dark matter halos are processed with the Santa Cruz semi-analytic model (SAM) for galaxy formation \citep{somerville99, somerville15b}, with dark matter halo merger trees constructed on-the-fly using an extended Press-Schechter (EPS)-based algorithm \citep{somerville99b}. We refer the reader to \citet{yung22} for detail regarding the construction of the simulated lightcones.

The Santa Cruz SAM tracks a wide variety of baryonic processes using prescriptions derived analytically, inferred by observations or extracted from numerical simulations, and provides physically-backed predictions for galaxies across wide ranges of redshift and mass. This model has been shown to be able to reproduce the observed evolution in distribution functions of rest-frame UV luminosity, stellar mass, and SFR from $z \sim 0$ to the highest redshift where observational constraints are available \citep{somerville15b, yung19, yung19b}.
The model performance during the epoch of reionization has been extensively tested and shown to agree extremely well with the observed evolution in one-point distribution functions of many galaxy properties, scaling relations, IGM and CMB reionization constraints, and two-point correlation functions \citep{yung19, yung19b, yung20b, yung20, yung21, yung22}. 

The physically-predicted properties and star-formation history (SFH) are assigned SEDs which are generated based on stellar population synthesis (SPS) models by \citet{bruzual03}. In addition, we include nebular emission lines predicted by numerical models from \citet{hirschmann17, hirschmann19}. These models account for excitation from young stellar populations, feedback from accreting supermassive black holes, and post-AGN stars. The nebular emission lines-included are forward-modelled into a rest-frame UV luminosity and observed-frame JWST photometry, including ISM dust attenuation \citep{calzetti00} and IGM extinction \citep{madau96}.

\begin{figure}[h]
    \centering
    \includegraphics[width=0.45\textwidth]{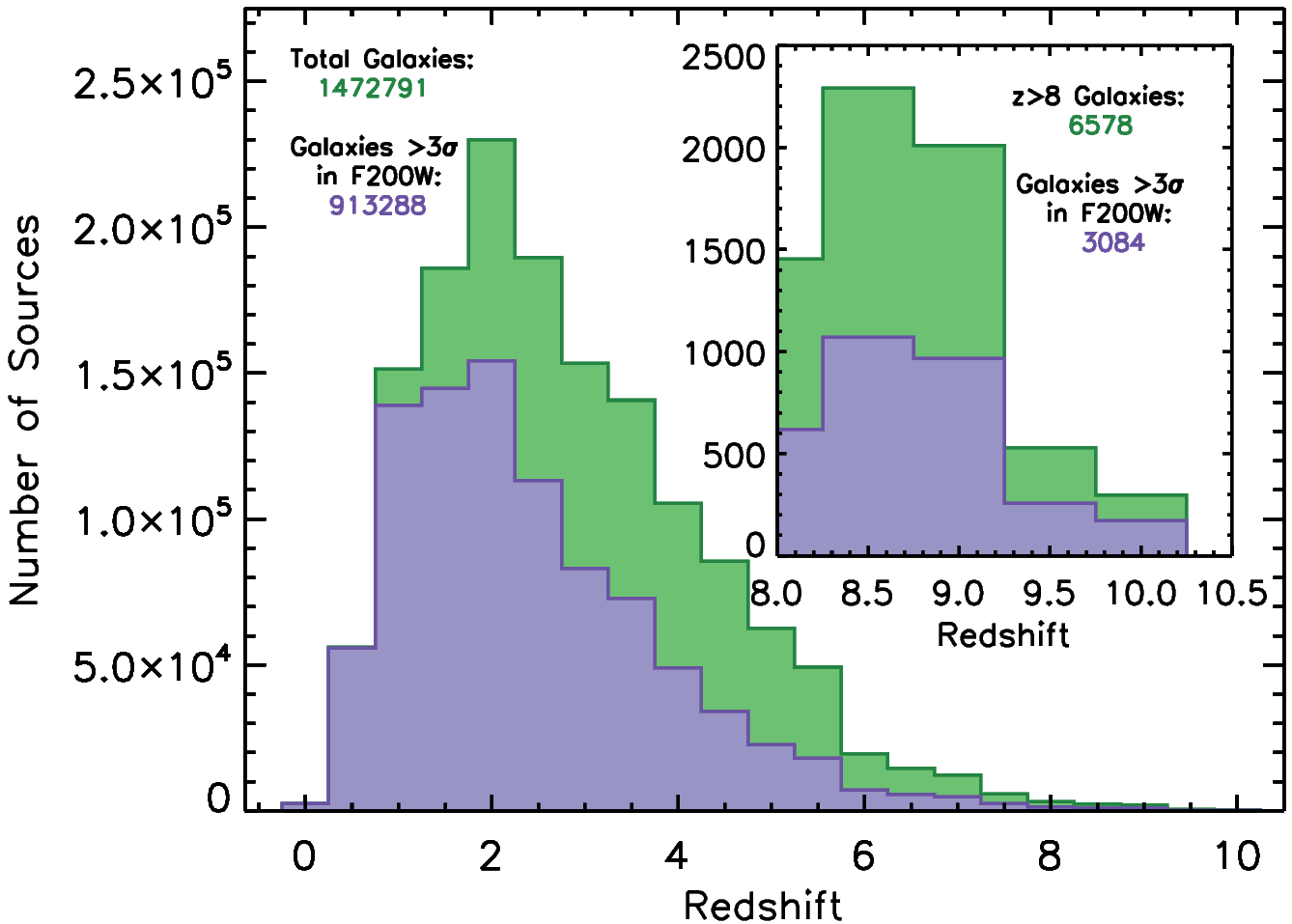}
    \caption{The redshift distribution of the full SAM catalog of 1,472,791 galaxies is shown in green with a zoom in on the 6,578 $z=8-10$ galaxies from the CEERS Simulated Data Product V3 \citep[][]{yung22}. As real observations with \jwst\ are limited by our ability to detect objects in the images we impose an initial S/N > 3 cut in F200W, where the noise is set to the 1$\sigma$ expected CEERS depth ($m_{3\sigma}=29.5$; \citealt{ceersprop}). This paper utilizes the 913,288 galaxies (3084 at $z>8$) that meet this initial criterion (purple). }
    \label{fig:redshiftdist}
\end{figure}

For this project we use the published CEERS Simulated Data Product V3\footnote{\href{https://ceers.github.io/releases.html}{ceers.github.io/releases}} catalog which includes the 0th realization of the SAM containing 1,472,791 total galaxies. The redshift distribution of the full SAM catalog is shown in Figure \ref{fig:redshiftdist} (green) with a zoom in on the 6,578 $z=8-10$ galaxies.  As real observations with \jwst\ are limited by our ability to detect objects in the images, we impose a S/N$>3$ cut in F200W where the ``noise" is set to the expected 1$\sigma$ CEERS depth (m=30.72; \citealt{ceersprop}). The rest of this paper utilizes the 913,288 galaxies (3084 at $z>8$) that meet this criterion (purple).

\subsection{Comparing Template Colors to Predicted z > 8 Galaxies}

To perform our photometric redshift estimations we use the photometric-redshift (photo-z) code EAZY \citep{brammer08}, and the included templates. The latest EAZY template set, known as “tweak\_fsps\_QSF\_v12\_v3” is based on the Flexible Stellar Population Synthesis (FSPS) code \citep{conroy10}. This template set has further been corrected (or “tweaked”) for systematic offsets observed between data and the models. \citet{finkelstein22a} found that the native EAZY FSPS templates were redder than their sample of observed $z=6-8$ galaxies. To cover a larger color-range which better represented their high-z sample, they added as an additional template the observed spectrum of the $z = 2.3$ galaxy BX418, which is young, low-mass, and blue \citep{erb10}. This galaxy’s color is 0.12 mag bluer than the bluest EAZY FSPS template, and has a color bluer than 85\% of the known high-redshift galaxies at the time. \citet{finkelstein22a} add two versions of this template; one with the observed \lya\ emission, and one where \lya\ was removed, to account for blue galaxies whose \lya\ has been absorbed from a potentially neutral IGM \citep[e.g.][]{miralda-escude98, malhotra06, dijkstra14}.

We first test whether the colors of the native EAZY FSPS templates, plus the single bluer \citet{erb10} template added by \citet{finkelstein22a}, cover the full color-space of our simulated galaxies. We redshifted these templates to to $z =$ 10 (a reasonable redshift of ``first discovery" for {\it JWST}), and measured the {\it JWST}/NIRCam F200W $-$ F277W color.  We chose this color as it measures the rest-frame UV color around 2000\AA\ at this redshift, and it is fully red-ward of the \lya\ break for $z \lesssim$ 13.5. Since we are only measuring a color between two filters we do not normalize the templates, and only multiply the wavelength by $(1+z)$. 

The templates included with the EAZY software are in \flam\ units, so we must first convert them into \fnu\ and then pass these templates through both the NIRCAM F200W and F277W filters\footnote{\href{https://jwst-docs.stsci.edu/jwst-near-infrared-camera/nircam-instrumentation/nircam-filters}{jwst-docs.stsci.edu/jwst-near-infrared-camera/nircam-instrumentation/nircam-filters}} by interpolating the filter transmission curve onto the SED template wavelength array. We then set any values for the filter transmission that are negative after interpolation or are smaller than 0.001 to 0.0 and integrate the SED template through the filter using 
$$
F_\nu (z,t,\tau,A_{1700},M) = \frac{\int T_\nu F_\nu (z,\lambda, t, \tau, A_{\lambda_0} ) \frac{d \nu}{\nu}}{\int T_\nu \frac{d \nu}{\nu}}
$$
where $T_\nu$ is the transmission curve for the filter, and $F_\nu$ is the flux of the SED template \citep{papovich01}. This gives the flux bandpass-averaged flux, $f_{band}$, in that filter band. We then measure the F200W $-$ F277W color of the template as 
$$ 
\mathrm{m_{F200W}-m_{F277W} = -2.5 log_{10}}\left(\frac{f_{\mathrm{F200W}}}{f_{\mathrm{F277W}}}\right) 
$$ 
where redder colors would have more positive values, and bluer colors would have more negative values. We list the F200W $-$ F277W colors of the native EAZY FSPS template set, and the additional \citet{erb10} template used by \citet{finkelstein22a} in Table \ref{tab:templatecolors}, noting that at the \jwst\ wavelengths the \citet{erb10} template is still bluer (more negative) than all of the EAZY FSPS templates. 

\begin{table}[h]
    \centering
    \begin{tabular}{c|c}
        \textbf{Template Name} &  \textbf{F200W $-$ F277W color} \\
        \hline
        tweak\_fsps\_QSF\_12\_v3\_001   &   0.302 \\
        tweak\_fsps\_QSF\_12\_v3\_002   &   0.783 \\
        tweak\_fsps\_QSF\_12\_v3\_003   &   1.611 \\
        tweak\_fsps\_QSF\_12\_v3\_004   &   1.744 \\
        tweak\_fsps\_QSF\_12\_v3\_005   &   1.787 \\
        tweak\_fsps\_QSF\_12\_v3\_006   &   1.055 \\
        tweak\_fsps\_QSF\_12\_v3\_007   &   -0.062 \\
        tweak\_fsps\_QSF\_12\_v3\_008   &   0.127 \\
        tweak\_fsps\_QSF\_12\_v3\_009   &   0.538 \\
        tweak\_fsps\_QSF\_12\_v3\_010   &   0.996 \\
        tweak\_fsps\_QSF\_12\_v3\_011   &   1.316 \\
        tweak\_fsps\_QSF\_12\_v3\_012   &   1.367 \\
        \hline
        erb2010\_highEW         &   -0.211 \\

        \textbf{New Template Name} &  \textbf{F200W $-$ F277W color} \\
        \hline
        binc100z001age6   &   -0.428 \\
        binc100z001age65  &   -0.343 \\
        binc100z001age7   &   -0.291 \\
        \hline
        binc100z001age6\_cloudy   &   -0.280 \\
        binc100z001age65\_cloudy  &   -0.259 \\
        binc100z001age7\_cloudy   &   -0.243 \\
    \end{tabular}
    \caption{SED template F200W $-$ F277W colors used for high-redshift galaxy target selection. Each of the templates has been redshifted to $z=10$ for this measurement. The ``tweak FSPS"  models are distributed with the EAZY software \citep[][]{brammer08}. A template based on the observations of \citet{erb10} had been previously included by \citet{finkelstein22a} for high-z galaxies in order to include a bluer template that matched the colors of their $z=6-8$ sample. We created BPASS and BPASS $+$ {\sc Cloudy} emission line templates to fully cover the color space of simulated high-redshift galaxies. We note that the nebular continuum emission included in the BPASS $+$ {\sc Cloudy} templates makes them redder in color than the BPASS only models that do not include emission lines.}
    \label{tab:templatecolors}
\end{table}

\begin{figure}[h!]
    \centering
    \includegraphics[width=0.48\textwidth]{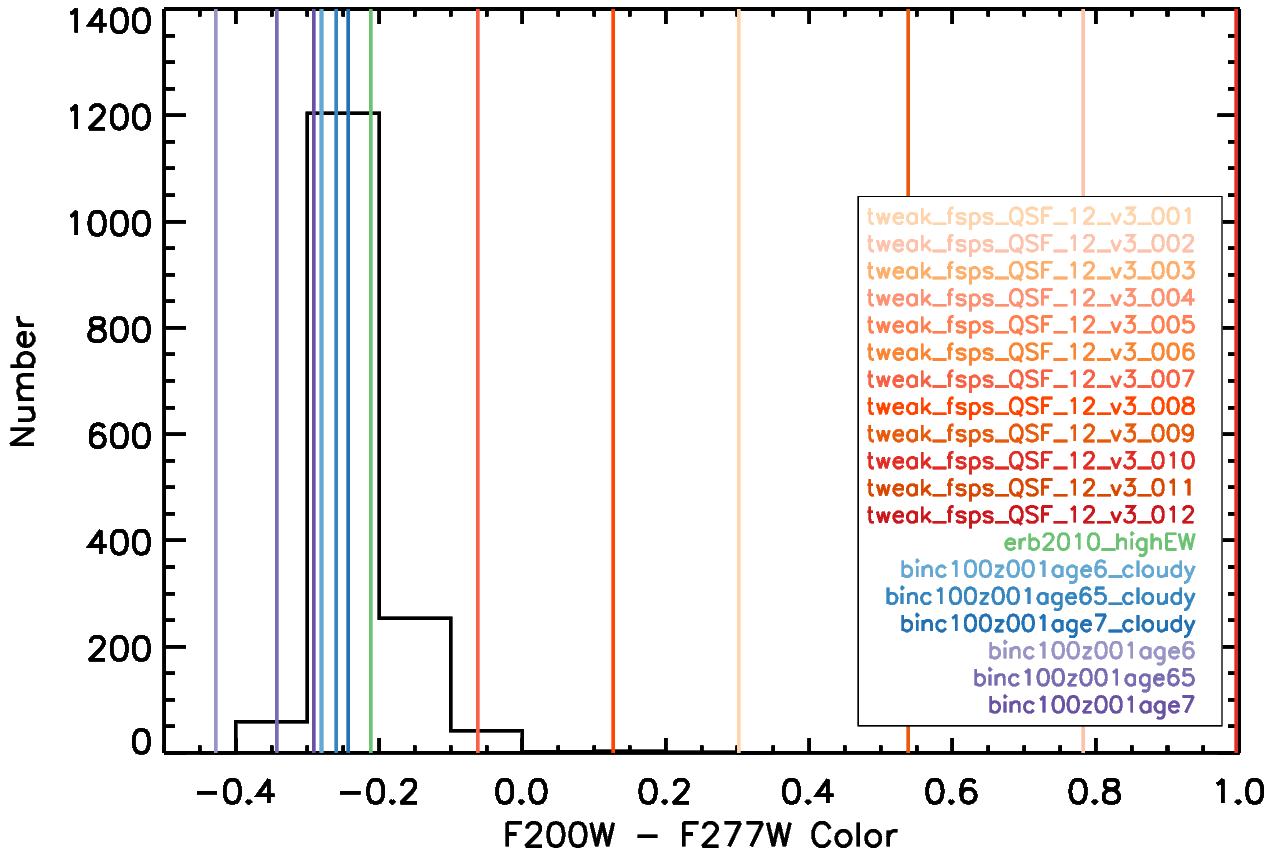}
    \caption{The black histogram shows the distribution of F200W $-$ F277W colors for the $z>8$ galaxies from the SAM catalog of \citet{yung22}. The solid vertical lines show the rest-UV color of the SED templates we used for photometric-redshift fitting, using the color calculated by integrating the templates through the \jwst/NIRCam F200W and F277W filters after placing them at $z=10$. The bluest EAZY FSPS template only reaches a rest-UV color of $-$0.1, while the majority of the comparison high-redshift sample have bluer (more negative) colors. \citet{finkelstein22a} added a bluer template from \citet{erb10} (green), but it is still redder than the majority of our simulated high-redshift galaxies. We created BPASS (purple) and BPASS $+$ {\sc Cloudy} emission line (blue) templates and note that the BPASS $+$ {\sc Cloudy} templates are redder in color than the BPASS only models due to their nebular continuum emission. This full template set can now reproduce the colors of all high-redshift galaxies in our simulated sample.}
    \label{fig:templatecolors}
\end{figure}

We compared the color space spanned by these templates (vertical lines) to our simulated galaxies (black histogram) from \citet{yung22} in Figure \ref{fig:templatecolors}. We find that the FSPS templates that are included with EAZY are all much redder than our simulated  $z>8$ galaxies. We also note that the template from \citet{erb10} that was added by \citet{finkelstein22a}, while bluer than the FSPS templates, is still redder than a majority of our simulated high-redshift galaxies. This shows that bluer models are needed in our template set to better represent the expected colors of $z >$ 8 galaxies and ensure accurate SED fits with \jwst\ data.

\section{Creating Blue Galaxy SED Templates}
As none of the EAZY FSPS templates have colors blue enough to match our simulated high-redshift ($z>8$) galaxies we created new, bluer templates that would more accurately represent our target galaxies.

\subsection{BPASS Templates}
 We created model SED templates using BPASS v2.2.1\footnote{\href{https://bpass.auckland.ac.nz/}{bpass.auckland.ac.nz/}} \citep{eldridge17, stanway18} which contain low metalicities (as expected in the high-redshift Universe), young stellar populations (since not much time has passed since the Big Bang at $z > 8$), and which also include binary stars. We chose the templates that used the \citet{chabrier03} 100 M$_{\odot}$ upper mass limit on the stellar initial mass function (IMF), and note that when we looked at the 300 M$_{\odot}$ mass-limit IMF templates the colors at $z=10$ did not change significantly. These BPASS templates do not include emission lines and all have a low metallicity, $Z=0.001$ (5\% $Z_{\odot}$). We created 3 templates named: binc100z001age6, binc100z001age65, and binc100z001age7 which have log stellar ages of 6, 6.5, and 7 Myr respectively. 
 
\begin{figure*}[ht!]
    \centering
    \includegraphics[width=0.95\textwidth]{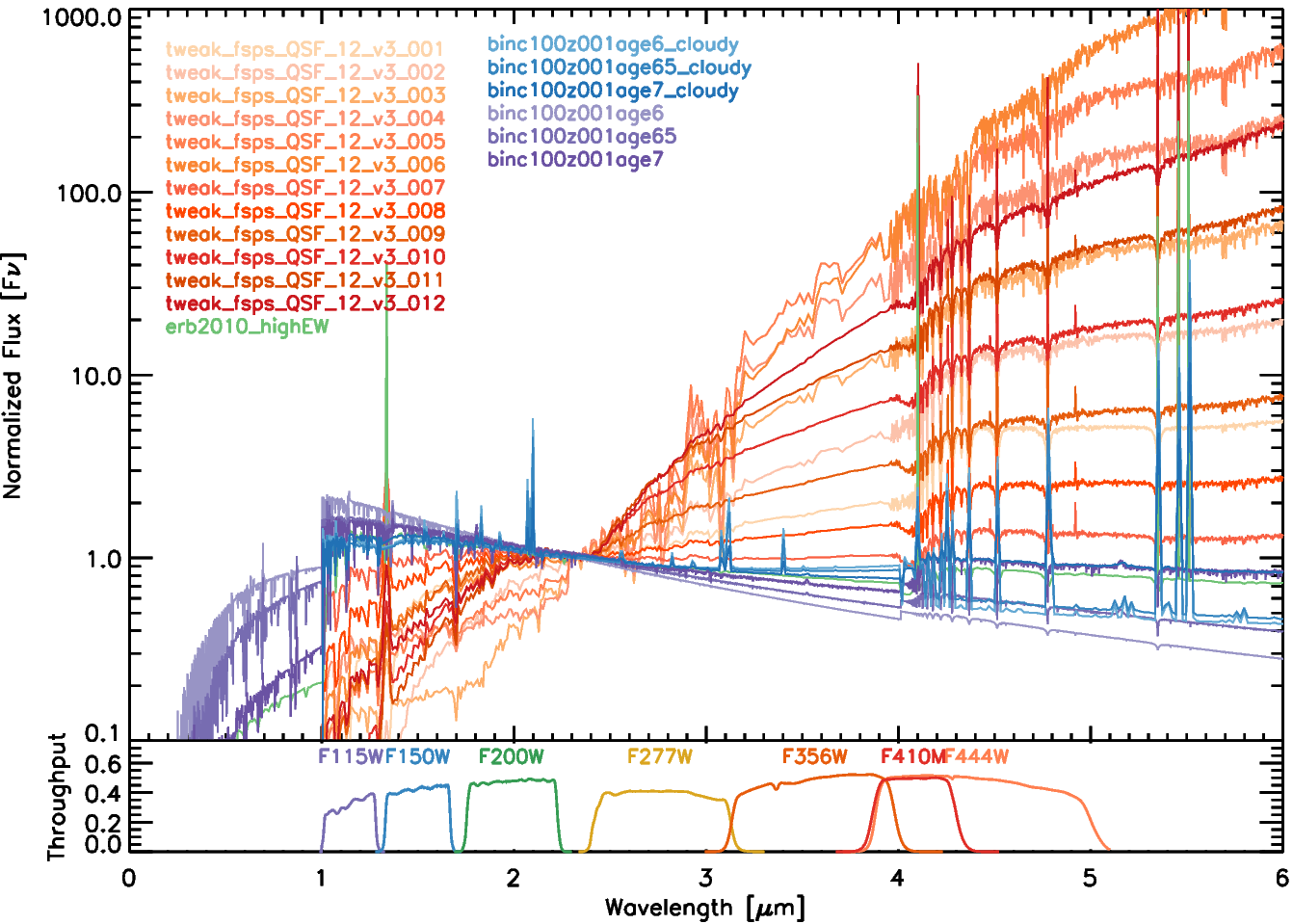}
    \caption{The rest-frame ultraviolet region of the EAZY template set, redshifted to $z=10$, used in our analysis to measure photometric redshifts. The red and orange lines show the latest standard EAZY template set (tweak\_fsps\_QSF\_v12\_v3), while the purple lines show the BPASS models we create here as described in \S 3.3. The blue lines show the BPASS + {\sc Cloudy} templates that have high nebular-line EWs, as described in \S 3.4. As can be seen, these newly created templates are bluer than the standard set, better matching the expected colors of $z >$ 8 galaxies.  All templates are normalized to their flux density at 2.301$\mu$m. We also include in this plot the template from \citet{erb10} used by \citet{finkelstein22a} which includes a high-EW$_{Ly\alpha}$ emission line. This template was not used in our analysis as the new BPASS and {\sc Cloudy} templates satisfied the same color range and our fits were not improved with its inclusion.}
    \label{fig:alltemplates}
\end{figure*}

The BPASS templates are in \AA\ and the flux is in L$_{\odot}$ (L$_\lambda$) so we must first convert them into F$_\lambda$ for EAZY. These BPASS models have high spectral resolution, thus we rebin them from $\Delta \lambda =$ 1\AA\ to $\Delta \lambda =$ 10\AA. We list the measured F200W\,$-$\,F277W colors of our new BPASS templates in Table \ref{tab:templatecolors} and plot them in purple in Figures \ref{fig:templatecolors} and \ref{fig:alltemplates}. The addition of these templates results in F200W\,$-$\,F277W colors to $<-0.4$, which is bluer than any of our $z>8$ galaxies in the SAM. This ensures that we are fully covering the color-space of our simulated galaxies, thus providing SED models which accurately match the data, resulting in more-accurate photometric redshifts, as we show in \S 4.

\subsection{{\sc Cloudy} Nebular Emission}

While the BPASS templates do not include any emission lines in their spectra, the simulated (and real) galaxies do, thus we explore adding emission lines to these new BPASS templates. Similar studies focused upon high-redshift galaxies have found that models with higher ionization parameters (log\,U $> -2.5$) and lower metallicities ($Z \lesssim 0.3\, Z_\odot$) better reproduce the observed properties \citep[e.g.][]{Inoue16, jaskot16, stark17, hutchison19, topping21}.  This effect has been seen with lower-redshift analogue samples as well \citep[e.g.][]{sobral18, berg16, berg18, berg19, tang19, tang21}.  In several instances (both in high-redshift and lower-redshift analogue sources), observations paired to photoionization modeling have suggested metallicities as low as $Z \sim 0.03-0.15\, Z_\odot$ \citep[e.g.][]{erb10,stark15, stark15b, vanzella16, berg18, berg21, senchyna21}, low values which we anticipate may be increasingly common the higher in redshift, and further back in time, we probe.

Motivated by these and other studies, we model the emission line spectra using {\sc Cloudy} v17.0 \citep{ferland17} with an ionization parameter log U $=-2$ and with the gas-phase metallicity = 0.05 $Z_{\odot}$ (fixed to stellar metallicity). In line with the prescription of other higher-redshift modeling \citep[e.g.][]{jaskot16, steidel16, stark16, stark17, hutchison19}, we set the Hydrogen density of the gas to be 300 cm$^{-3}$, assume a spherical geometry for the nebular gas, and set the covering factor of the gas to be 100\%. 

As the high-redshift galaxies we are specifically targeting are likely to have little to no detectable \lya\ emission due to attenuation by the neutral IGM during this epoch, we removed this emission feature from the template spectra. We also note that the SAM galaxies do not include \lya\ emission, thus by doing this we are choosing templates that more accurately represent our simulated data. To remove the \lya\ emission feature, we cut out the array between 0.120 and 0.125 $\mu$m and interpolate a flat continuum line over that range.

We note that {\sc Cloudy} creates both nebular line  and nebular continuum emission; the latter results in a moderate reddening of the continuum slope. The BPASS$+${\sc Cloudy} models are redder in  the F200W $-$ F277W color than the BPASS-only models due to this effect. Our 3 additional BPASS$+${\sc Cloudy} templates are named: binc100z001age6\_cloudy, binc100z001age65\_cloudy, and binc100z001age7\_cloudy. We list the measured F200W - F277W colors of these new template in Table \ref{tab:templatecolors} and plot them in blue in both Figures \ref{fig:templatecolors} and \ref{fig:alltemplates}.

\begin{figure*}
    \centering
    \includegraphics[width=0.33\textwidth]{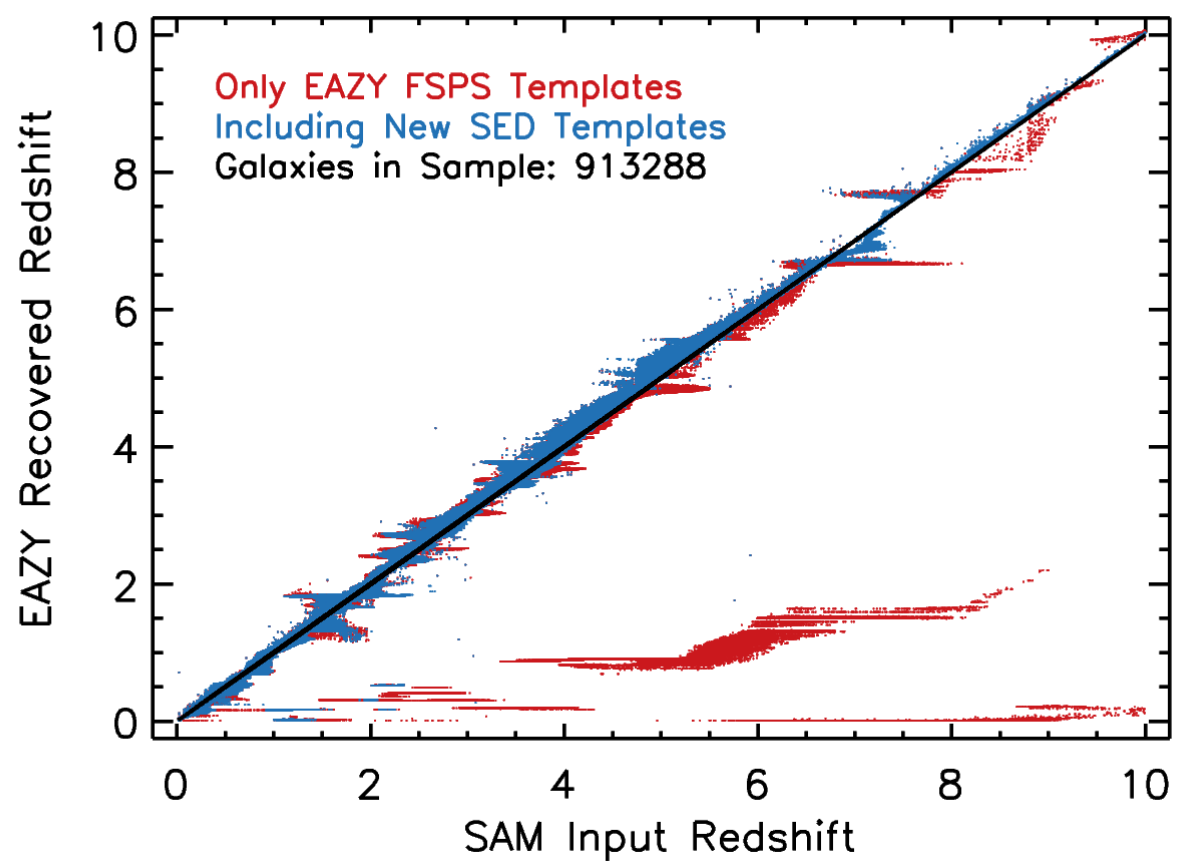}
    \includegraphics[width=0.33\textwidth]{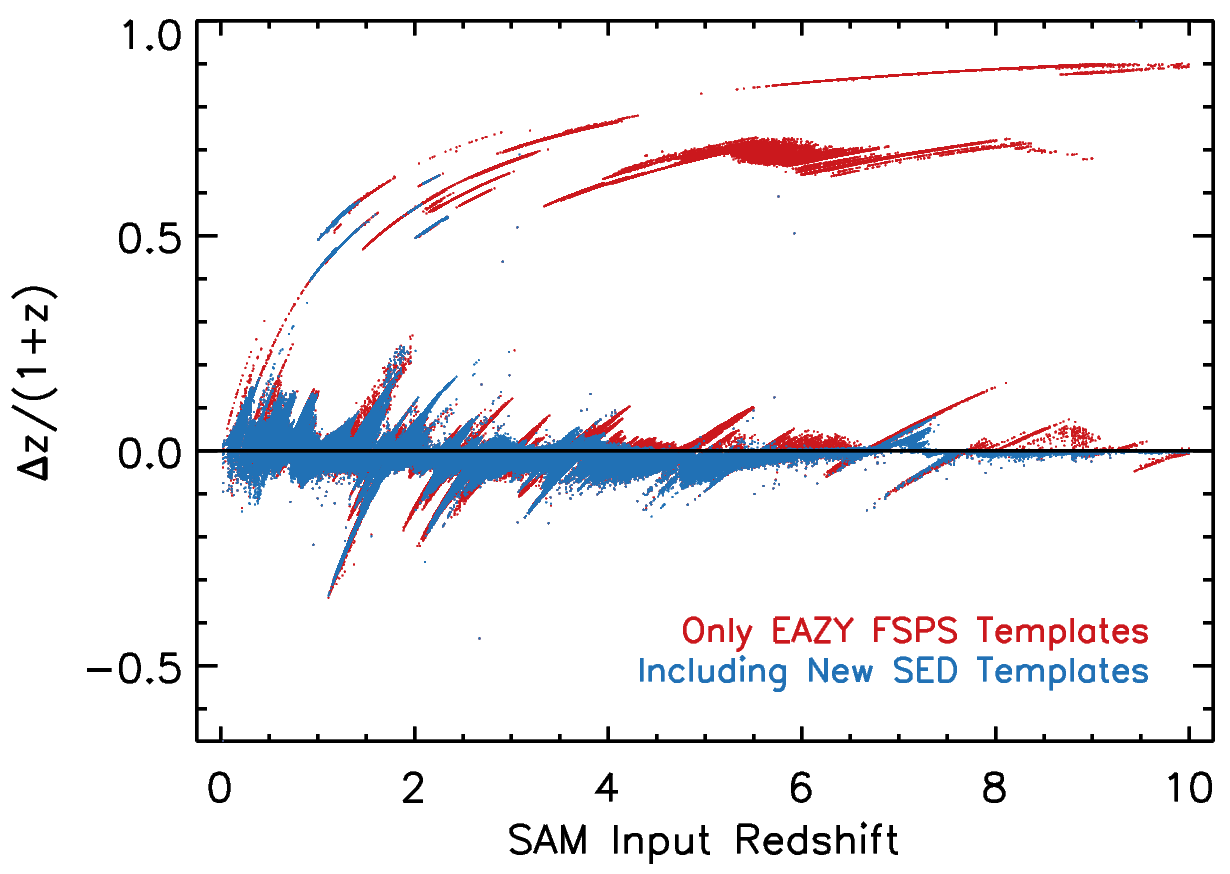}
    \includegraphics[width=0.33\textwidth]{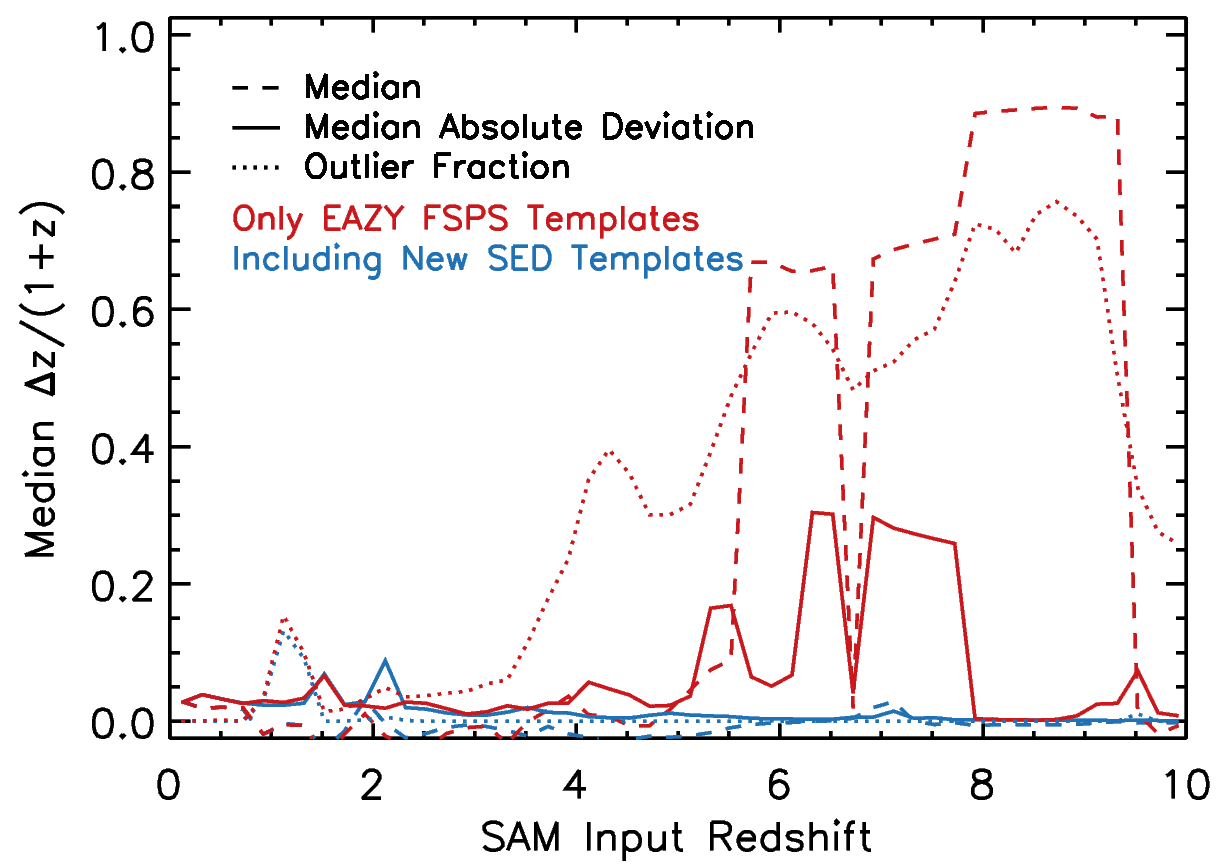}
    \caption{{\textbf Left:} Comparison of our recovered redshifts from EAZY vs the input redshifts from the SAM. We ran our sample of 913288 simulated galaxies from $z=0-10$ through EAZY using only the included EAZY FSPS templates (red), and then again after adding our new SED template set (blue) as described in \S 3.3 and Figure \ref{fig:alltemplates}. For this test we set the errors in each filter equivalent to a 5$\sigma$ depth of $m=30$. {\textbf Center:} The $\Delta z$ (input redshift - recovered redshift) vs input redshifts of our SAM galaxies from both EAZY runs. {\textbf Right:} We show the median (dashed line), and median absolute  deviation (solid line) of the $\Delta z$ for both EAZY runs, where the inclusion of the new templates provides significant improvement as both values are lower across the full redshift range. We calculate and outlier fraction, or catastrophic failures, as those where $\Delta z > 0.2 \times z$ (dotted line) which highlights the set of $z>4$ galaxies that are fit at lower redshifts when using only the original templates, but whose redshifts are accurately recovered after the inclusion of our new set of SED templates.}
    \label{fig:newvoldz}
\end{figure*}

\subsection{New Suite of Blue SED Templates for Use with EAZY}

To ensure that we are covering the color space of our high-redshift galaxies we compare the F200W - F277W color distribution of the $z>8$ galaxies in the SAM to the colors of our full template set in Figure \ref{fig:templatecolors}. The distribution of SAM F200W - F277W galaxy colors are shown by the black histogram in Figure \ref{fig:templatecolors} while the color for each SED template is plotted as a vertical line. With the addition of our six new templates (three with and three without {\sc Cloudy} nebular emission) we now have a set of SED templates that represent the full range of rest-UV colors of our simulated $z>8$ galaxies. We report the colors for each template in Table \ref{tab:templatecolors}. The full set of templates redshifted to $z=10$, inclusive of our new BPASS and BPASS$+${\sc Cloudy} templates, are plotted in Figure \ref{fig:alltemplates}. The plot is normalized to the flux at 2.3$\mu$m as this is between the F200W and F277W filters and shows the slope (color) between them visually.

We note that by including templates that have the {\sc Cloudy} parameters detailed above, we are assuming an escape fraction, $f_{\mathrm esc}=0$. This may not be true of our high-redshift ($z>8$) galaxies, but since we include both the BPASS templates without nebular emission, and the BPASS$+${\sc Cloudy} templates with it, EAZY's linear combination of templates can generate a composite SED for any level of escape fraction.

For this project we used the new template set described above where \lya\ has been removed from the spectra as the IGM attenuation at high redshifts ($z>8$) impacts its transmission and is not included in the SAM galaxies. We make these new SED templates public for the community to utilize and provide sets of them without \lya\ (for high-redshift galaxies), with reduced \lya\ emission (either 1/3 or 1/10 of that produced by {\sc Cloudy}), and with full \lya\ strength. The templates, corresponding EAZY parameter files, and descriptions can be found at \href{https://ceers.github.io/LarsonSEDTemplates}{ceers.github.io/LarsonSEDTemplates}.

\section{Improvements to Photometric Redshift Fits with these New Templates}

With the new set of SED templates which spans the full color range of our simulated galaxies we tested if the addition of these templates improves our photometric redshift fits. After making the 3$\sigma$ cut in F200W on the SAM as described in \S 2 we have 913288 galaxies ranging from $z=0-10$ that we run through EAZY, using the 7 NIRCam filters from the CEERS Survey (F115W, F150W, F200W, F277W, F356W, F410M, and F444W) plus 4 \hst\ filters from CANDELS (ACS F606W, ACS F814W, WFC3 F125W, and WFC3 F160W) to fit photometric redshifts. We allowed the redshift to span from $0.1 < z < 15$, in steps of 0.01, and adopted a flat luminosity prior as we are just beginning to explore galaxies at early times. For our reported recovered redshift we use the output $z_a$ value from EAZY.

\subsection{Testing Redshift Recovery with Updated Template Set}
To determine if our additional, bluer templates improve the redshift fits we compared the recovered redshift from EAZY to the input redshift from the SAM with and without the inclusion of our additional templates, setting the flux uncertainties as equivalent to a 5$\sigma$ $m=30$ depth in each filter (e.g., 1$\sigma$ noise of 0.73 nJy), and without perturbing the SAM fluxes for this test. We did two runs of EAZY on the full catalog, first using only the original FSPS templates, and then again using the full set of SED templates (FSPS, BPASS, and BPASS$+${\sc Cloudy}). Figure \ref{fig:newvoldz} (Left) shows the recovered redshifts from EAZY compared to the input redshifts from the SAM for both the run using the EAZY FSPS templates (red) and after including our new templates (blue). There is significant improvement in recovering the correct redshifts with the new templates, as with the old templates the fits chose $z=0 \sim 2$ solutions for many of the $z>2$ galaxies. We also show the difference between recovered and input redshift ($\Delta z$) for all of our photometric redshift fits as a function of input redshift in Figure \ref{fig:newvoldz} (middle) where the accuracy of our recovered redshifts using the new templates (blue) is much higher than just with the original EAZY templates (red), especially at the high redshifts of interest ($z>8$). 

\begin{figure*}[ht!]
    \centering
    \includegraphics[width=0.95\textwidth]{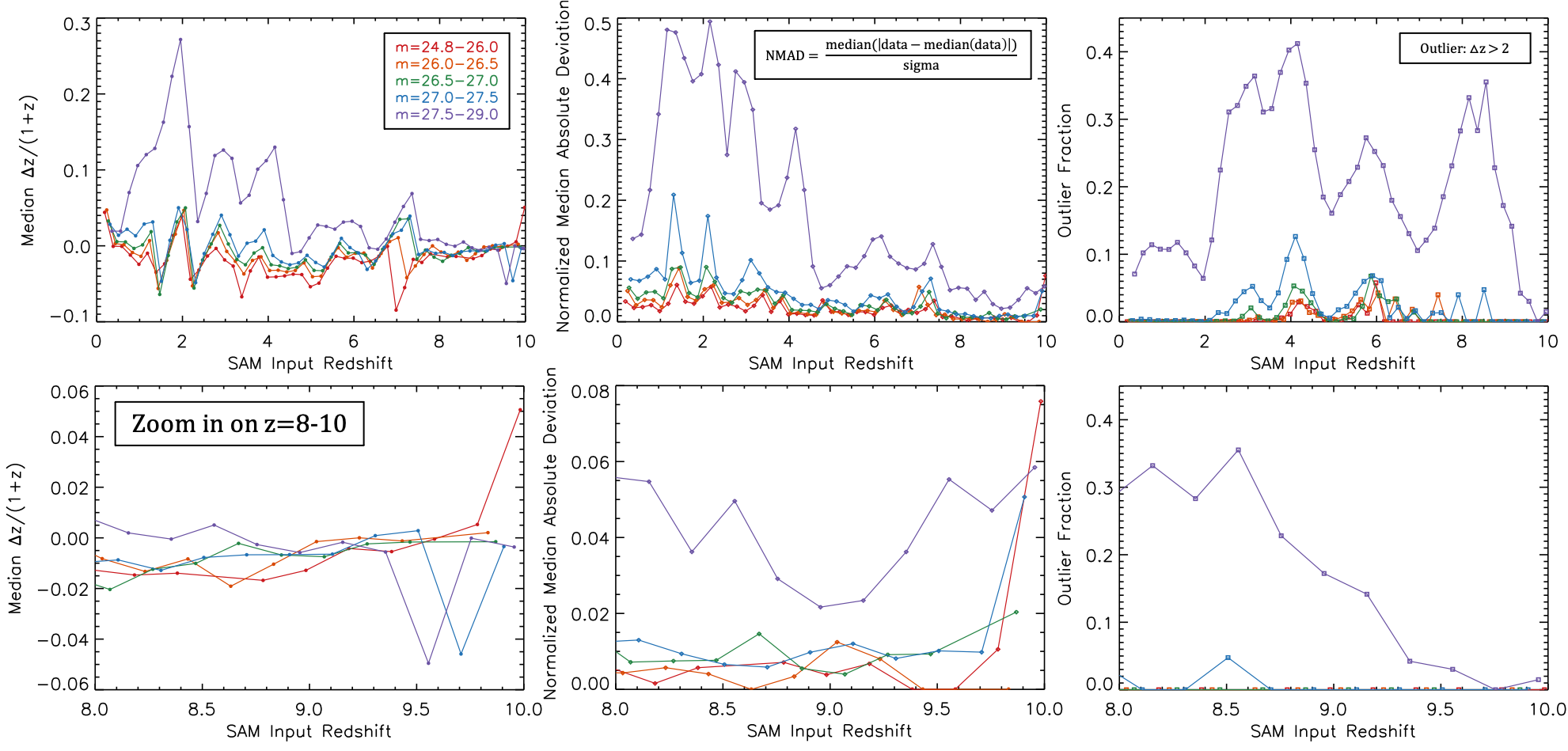}
    \caption{Plots illustrating the accuracy of our recovered redshifts from EAZY versus input redshifts for our mock CEERS observations of the SAM galaxies. We show the full $z=0-10$ set of 913,288 galaxies in the SAM ({\textbf Top}), and a zoom in on the $z=8-10$ range of particular interest to \jwst\ studies ({\textbf Bottom}). {\textbf Left:} The median $\Delta z$ as a function of redshift, separated in bins of F200W magnitude. {\textbf Center:} The standard deviation (NMAD) of our $\Delta z$ in corresponding magnitude bins. {\textbf Right:} The fraction of outliers in our fits where we define outliers as those with a $\Delta z >2$ binned in magnitude. For each of these parameters we note that the faint galaxies ($m>27.5$, purple lines) are the sources with the least-accurate recovered redshifts, which is not unexpected as constraints on their colors are the poorest.}
    \label{fig:medianstddev}
\end{figure*}

To measure how well we recover the input redshifts with EAZY we calculate the median and standard deviation of $\frac{\Delta z}{1+z}$ in bins of $\Delta z=0.2$ and magnitude bins of $\Delta m=0.5$ where we use a Normalized Median Absolute Deviation `NMAD' for the standard deviation calculations as it is more outlier-resistant. 
$$ \mathrm{NMAD} = \frac{\mathrm{median(|data - median(data)|)}}{\mathrm{sigma}}$$
Here sigma is the inverse of the error function of $0.5 \times \sqrt{2}$ or 0.67449 (assuming a Gaussian error distribution). We show the median (solid line) and NMAD (dashed line) of our photometric redshift fits both with (blue) and without (red) the new templates in Figure \ref{fig:newvoldz}. We also calculate an outlier fraction where we define outliers as those where $\Delta z > 0.2 \times z$ (dotted line). This highlights the improvement generated when including our new templates over the original EAZY template set. Across the redshift range of our SAM galaxies ($z=0-10$) our fits using the new templates do a significantly better job at accurately recovering the redshift of our sources than when using only the original EAZY templates; where the median $\Delta z$, NMAD, and outlier fraction are all lower.

\section{Creating a Simulated Catalog at CEERS Depths}

There are many upcoming surveys with {\it JWST} that will be searching for distant galaxies; one of which is the Cosmic Evolution Early Release Science (CEERS) Survey \citep[][]{bagley22} which covers 96.8 arcmin$^2$ to an expected 5$\sigma$ depth of m$\sim$28.6 \citep{ceersprop}.  CEERS has published simulated catalogs of the field and created mock observations using the lightcones from \citet{yung22}. Table \ref{tab:ceersdepths} shows the  expected $5\sigma$ depths in each of the CEERS filters\footnote{\href{https://ceers.github.io/obs.html}{ceers.github.io/obs}}, and includes the current \hst\ ACS and WFC3 depth over the same area from the CANDELS survey \citep{grogin11} as measured by \citet{finkelstein22a}. For all of the following tests we run EAZY using 4 HST filters: F606W, F814W, F125W, F160W, and 7 NIRCAM filters: F115W, F150W, F200W, F277W, F356W, F444W, and F410M. We use the 18 template set that includes the 12 tweak FSPS models, the 3 new BPASS templates described in \S 3.3, and the 3 new BPASS + {\sc Cloudy} templates described in \S 3.4. For the error in each filter we use the expected $1\sigma$ CEERS depth \citep[][]{ceersprop} for every galaxy and perturb the input fluxes of the sources to mimic expected errors in real data as described below.

\subsection{Perturbing Fluxes by Realistic Errors}

To best recreate realistic values for our simulated mock galaxies, we ``observe" their simulated fluxes by randomly perturbing them by an amount proportional to the expected flux errors. We do this via the method described in \citet{bagley22} where they modeled the noise to have a Voigt profile distribution (a Gaussian core with Lorentzian wings). We use the 1$\sigma$-depth in each filter (Table \ref{tab:ceersdepths}) as the Gaussian $\sigma$ for our perturbations. The following results use these perturbed fluxes for each of the simulated galaxies, with the $1\sigma$ depth as the error for each filter. 

~\\
\subsection{Measuring Accuracy of Photometric Redshifts with a Simulated JWST Catalog of Galaxies}

We run EAZY in the same manner as described above (with both the \hst\ and \jwst\ filters covering the CEERS field) on the 913,288 galaxies in the SAM using the perturbed fluxes and the CEERS 1$\sigma$ depth as the errors. For these runs we use the full template set that includes the original FSPS templates and our set of six new bluer ones. The goal is to determine how well we are able to recover the redshifts of our sources given our best approximation of true observing conditions. 

 To better quantify the accuracy of our redshift fits we show the median (left), NMAD (center), and outlier fraction (right) for our recovered redshifts from EAZY at the CEERS depths (Figure \ref{fig:medianstddev}). Here we define outliers as those with a $\Delta z >2$ binned in magnitude. We separate our measurements for each in magnitude bins as our ability to accurately fit and recover redshifts for real galaxies is magnitude dependent. For our magnitude distribution we use the perturbed F200W magnitudes.  This figure highlights that the recovered photometric redshifts are accurate across $z=$ 0--10 for sources with $m <$ 27.5 ($\approx$10$\sigma$ detections).  The accuracy progressively worsens for fainter galaxies, which is not unexpected as they are the hardest to detect and measure accurately.  However, as highlighted in the bottom row, even faint galaxies are fairly recoverable at $z >$ 9 as the Lyman break passes out of the deep CEERS F115W band, providing another dropout detection.

\begin{table}[h]
    \centering
\begin{tabular}{c|c|c|c|c}
     Filter  & 5$\sigma$ Depth  & 1$\sigma$ Error &  5$\sigma$ Depth &  1$\sigma$ Error  \\
      & (mag) & (mag) &  F$_\nu$ (nJy) &  F$_\nu$ (nJy) \\
     \hline
     ACS F606W  &   27.95  &   29.70 & 24.0 & 4.80 \\
     ACS F814W  &   27.60  &   29.35 & 33.1 & 6.62 \\
    WFC3 F125W  &   27.05  &   28.80 & 55.0 & 11.0 \\
    WFC3 F160W  &   27.10  &   28.85 & 52.5 & 10.5 \\
  NIRCAM F115W  &   29.15  &   30.90 & 7.94 & 1.59 \\
  NIRCAM F150W  &   28.90  &   30.65 & 10.0 & 2.00 \\
  NIRCAM F200W  &   28.97  &   30.72 & 9.38 & 1.88 \\
  NIRCAM F277W  &   29.15  &   30.90 & 7.94 & 1.59 \\
  NIRCAM F356W  &   28.95  &   30.70 & 9.55 & 1.91 \\
  NIRCAM F444W  &   28.60  &   30.35 & 13.2 & 2.64 \\
  NIRCAM F410M  &   28.40  &   30.15 & 15.9 & 3.17 \\
\end{tabular}
    \caption{The reported CANDELS \hst\ \citep[][]{grogin11} and expected CEERS \citep[][]{ceersprop} \jwst/NIRCam $5\sigma$ depths and the $1 \sigma$ errors used to perturb the SAM galaxies, shown in both magnitude and flux ($F_\nu$). }
    \label{tab:ceersdepths}
    \vspace{-7mm}
\end{table}

\section{Selection Criteria for z>8 Galaxies with JWST}

Through rigorous testing and analysis we detail below the selection criteria that best identifies robust high-redshift ($z>8$) galaxy samples with our simulated \jwst\ catalog, while minimizing the level of low-redshift ($z<5$) interlopers in our sample. As above, we use EAZY to calculate redshifts probabilities, P(z)'s and we focus on $z>8$ as this epoch is made more accessible with \jwst's infrared wavelength coverage. At this redshift, galaxies drop out of the WFC3 F814W band due to their Lyman break at a rest frame of 1200 \AA, leaving only two \hst-band detections.  The CEERS \jwst\ filters reach further into the infrared, providing a wider range of photometric coverage for these galaxies, improving our ability to detect and characterize them, though we note that the \hst\ F814W data is crucial to probe the \lya\ break at $z \lesssim$ 9.\\

~\vspace{-5.4mm}\\
\noindent \textbf{S/N F200W \& S/N F277W > 5}: The first cut that we make on our catalog is in S/N in two of our filters, F200W and F277W. Requiring a significant detection in both the F200W and F277W bands aims to mimic detection bands in actual photometry, as both of these filters are red-ward of the \lya\ break in galaxies at these redshifts and should thus be detected by JWST at these wavelengths. We note that we ran tests on different S/N cuts in these bands individually and combined and find that $5\sigma$ in both removes a fair number of low-redshift ($z<5$) sources from the sample while not reducing the number of actual $z>8$ galaxies we recover. \\

~\vspace{-5.4mm}\\    
\noindent \textbf{$\int$P( z > 7) > 0.85}: The second cut we apply is one that requires $>85\%$ of the redshift P(z) to reside at $z > 8$ (integrated out to the maximum redshift we considered with EAZY of 15), allowing only $<15\%$ to be present in a low-redshift solution. Making this cut removes any flat P(z)'s, where the redshift is not well constrained by the SED fits, and any that might have significant peaks at low redshift, creating a robust sample of galaxies that are expected to be at high-redshift. We note that we also tried cuts at $z>8$, and ones that had higher percentages of their P(z) above the redshift cut (i.e. $<90\%$) or lower (i.e. $<75\%$), but our adopted criteria maximized our recovery rate of high-redshift galaxies while minimizing the contamination by low-redshift ($z<5$) galaxies. \\

\begin{figure*}[ht!]
    \centering
    \includegraphics[width=0.95\textwidth]{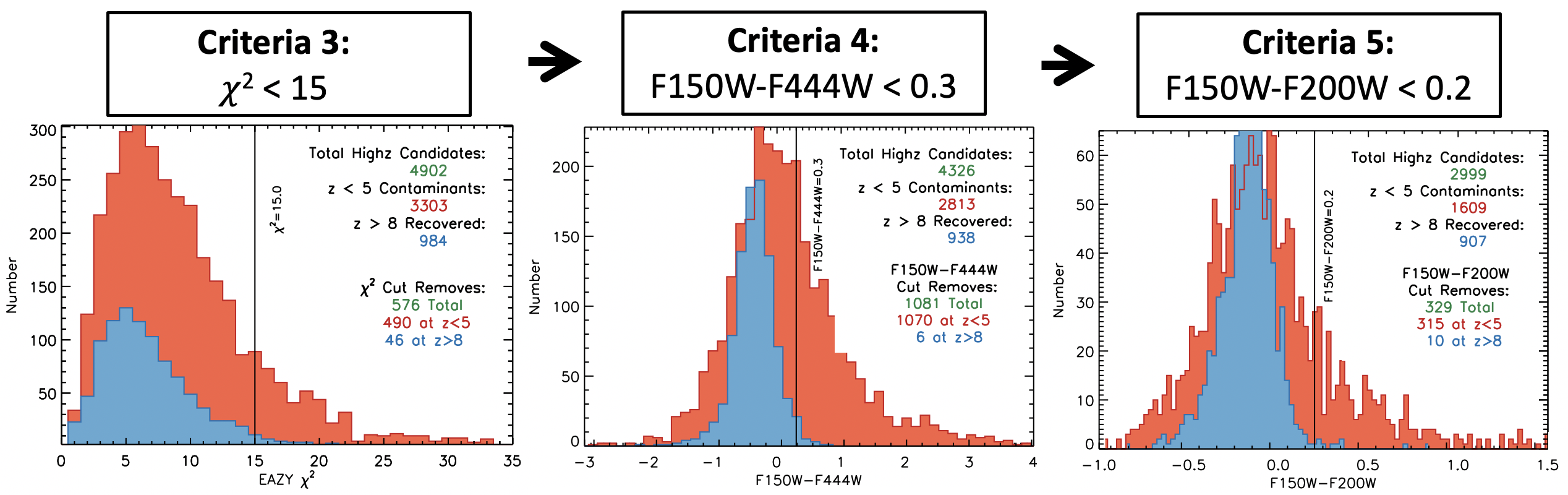}
    \caption{Plots illustrating the details of the selection criteria cuts we made to select for high-redshift galaxies and the impact each cut has on our final sample. In each figure we plot the real $z>8$ sources in our sample in blue and the contaminating $z<5$ galaxies in red. Prior to the cuts shown here we required a S/N $>5$ in both F200W and F277W and an $\int P( z > 7) > 0.85$ (see \S 6) which left us with a sample of 4902 potential high-redshift ($z>8$) galaxies.  We then make a cut in $\chi^2$ of the EAZY fit (\textbf{Left}) which removes 576 sources, 460 of which were contaminating low-redshift galaxies. In \S 6.1 we discuss the need for additional selection criteria that will remove contaminating low-redshift interlopers from our sample and that the above two color cuts: F150W $-$ F444W (\textbf{Center}) and F150W $-$ F200W (\textbf{Right}) were the two that most distinguished between these two sets of galaxies remaining in our sample. After requiring a F150W $-$ F444W color $< 0.3$ and F150W $-$ F200W $<0.2$ we are able to remove an additional 1410 sources, 1394 of which are contaminants, while only losing 16 real $z>8$ galaxies from the sample. We show the remaining 2670 sources in our sample and their distribution in true redshift and F200W magnitude in Figure \ref{fig:finalsample}, showing that a majority of the contaminants are at $m>28$. }
    \label{fig:criteria}
\end{figure*}

~\vspace{-5.4mm}\\
\noindent \textbf{$\chi^2$ < 15}: The third cut we require is for EAZY to have found a good fit to the data, rejecting objects where even the best-fitting solution is not a match to the observed photometry. The maximum allowed $\chi^2$ of the EAZY fit was also set to other values ranging from 15-35, but 15 was the best threshold we found for maximizing our recovery and minimizing our contamination rates of low-redshift galaxies. In the left-most panel of Figure \ref{fig:criteria} we show this distribution from our sample of galaxies, where the total number of high-redshift galaxies in our sample after making our first two cuts is 4902, with 3303 of those being contaminants (red, $z<5$) and 984 of those being actual $z>8$ galaxies (blue). Making this cut in $\chi^2$ removes 576 total sources, 490 being low-redshift contaminants while only removing 46 of our high-redshift ($z>8$) galaxies from our sample, leaving a remaining sample of 4326 candidate galaxies. 

\subsection{Color Cuts to Reduce Contamination}
Our sample of 4326 sources after these first three selection cuts still contains many low-redshift ($z<5$) galaxies (2813, $65\%$ of our sample) and thus we explored additional selection criteria that could differentiate between these contaminating sources and our actual high-redshift ($z>8$) galaxies in the SAM. Of the different criteria we explored, we found two color cuts which led to a direct distinction between the low- and high-redshift galaxies remaining in our sample. \\

\noindent \textbf{F150W $-$ F444W < 0.3}: The first color cut that we make requires F150W $-$ F444W $<0.3$, as this was the most distinct difference between our low- and high-redshift galaxies still remaining in our sample (see Figure \ref{fig:criteria} center panel). This cut removed 1081 galaxies from our sample of 4326 with only 6 of those being actual $z>8$ galaxies. This dropped our contamination rate from $65\%$ (where 2813/4326 galaxies were low-redshift) to $54\%$. This particular color spans a wide-range of wavelengths for these galaxies and in our SAM more of the low-redshift galaxies have a redder color where F444W is brighter than F150W, while the high-redshift galaxies are bluer and thus have a smaller/negative value. This was also evidenced by our measurement of the galaxy colors and the motivation for creating bluer templates for EAZY for this project (see \S 3.3). \\

\noindent \textbf{F150W $-$ F200W < 0.2}: The second color cut that we make is to require that F150W $-$ F200W $<0.2$ as shown in the right panel of Figure \ref{fig:criteria}. This cut removes an additional 329 galaxies from our sample with 315 of those being contaminating low-redshift ($z<5$) galaxies (red). This drops our contamination rate from $54\%$ to $48\%$ while still only removing 10 of our detectable input high-redshift ($z>8$) galaxies (blue). \\
 
 \begin{figure*}[ht!]
    \centering
     \includegraphics[width=0.8\textwidth]{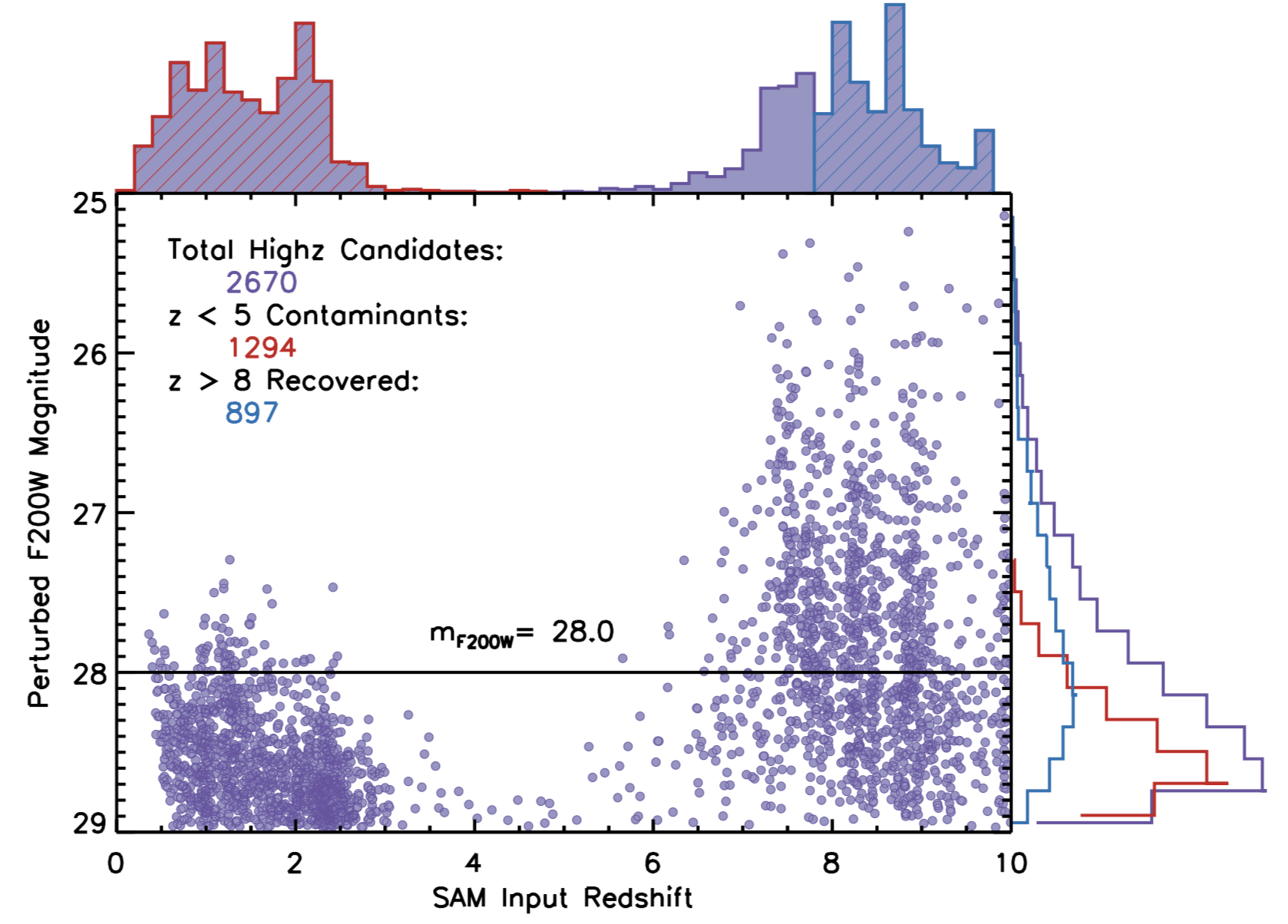}
    \caption{Our final sample of high-redshift galaxies after making cuts based upon the five selection detailed in \S 6. Here we plot these sources by their input redshift from the SAM vs the F200W perturbed magnitude with corresponding histograms for each axis. Our final sample includes 2670 sources, 1294 of which are contaminating $z<5$ galaxies while 897 are actual $z>8$ sources. As shown by the horizontal line, the sample is dominated by low-z contaminants at $m \gtrsim$ 28, but brighter than m$\sim$27 all of our selected sources are high-redshift galaxies. This shows that, with the colors from sources in this simulated catalog, near our survey detection limits we struggle with distinguishing these high-redshift sources from low-redshift interlopers, though we expect the true contamination rates in observations to be lower than those predicted here (\S 6.3). Details about our contamination and completeness fractions are shown in Figure \ref{fig:completecontam}. }
    \label{fig:finalsample}
\end{figure*}
 
After the above five selection criteria we are left with a remaining sample of 2670 galaxies, with 897 of those being real high-redshift ($z>8$) sources, and 1294 being low-redshift galaxies as shown in Figure \ref{fig:finalsample}. We note that many of our low-redshift contaminating sources are at $z<3$ and we plot them both as a function of input redshift from the SAM and F200W perturbed magnitude in Figure \ref{fig:finalsample}. Many of these contaminating galaxies are faint, $\mathrm{m_{F200W}>28}$ (horizontal line), which is not unexpected as the faint galaxies are harder to measure accurate colors for when their fluxes are closer to the detection limit. We also show the full distribution of input redshifts for our sample as histograms along the top axis of Figure \ref{fig:finalsample} marking in red the same galaxies we have been calling contaminants ($z<5$), and in blue those that we have designated as actual $z>8$ galaxies. 

\subsection{Calculating Completeness}
Here we define completeness as the number of detectable, real, $z>8$ sources that are recovered by EAZY as being at $z>8$, compared to the known number of true $z >$ 8 sources in the catalog. We define detectable as the sources in the lightcone that have input redshifts from the SAM above 8 and which also meet the S/N requirement of our selection criteria (here we use S/N F200W \& F277W > 5). In our SAM we have 6578 total galaxies at $z>8$, 3084 of which meet our initial cut at S/N $>3$ in F200W and run through EAZY (see inset in Figure \ref{fig:redshiftdist}). Here we only include those $z>8$ galaxies that meet the first sample cut of S/N $>5$ in both F200W and F277W as being truly detectable high-redshift sources, of which there are 1375. Of these sources in our final sample of high-redshift galaxies that meet all 5 of the selection criteria we recover 897 of the 1375, or  $65.2\%$. In Figure \ref{fig:completecontam} we show our completeness versus redshift, showing different magnitudes as the solid lines. We have higher completeness fractions above $z>9.5$ as at this redshift we gain a full dropout band with \jwst, F115W, significantly improving our SED fits as they have a distinctly detectable \lya\ break. We also suffer low completeness at the faint end of our high-redshift sources, where we are also dominated by contamination (see \S 6.3).

\subsection{Calculating Contamination}
\begin{figure}[ht]
    \centering
     \includegraphics[width=0.48\textwidth]{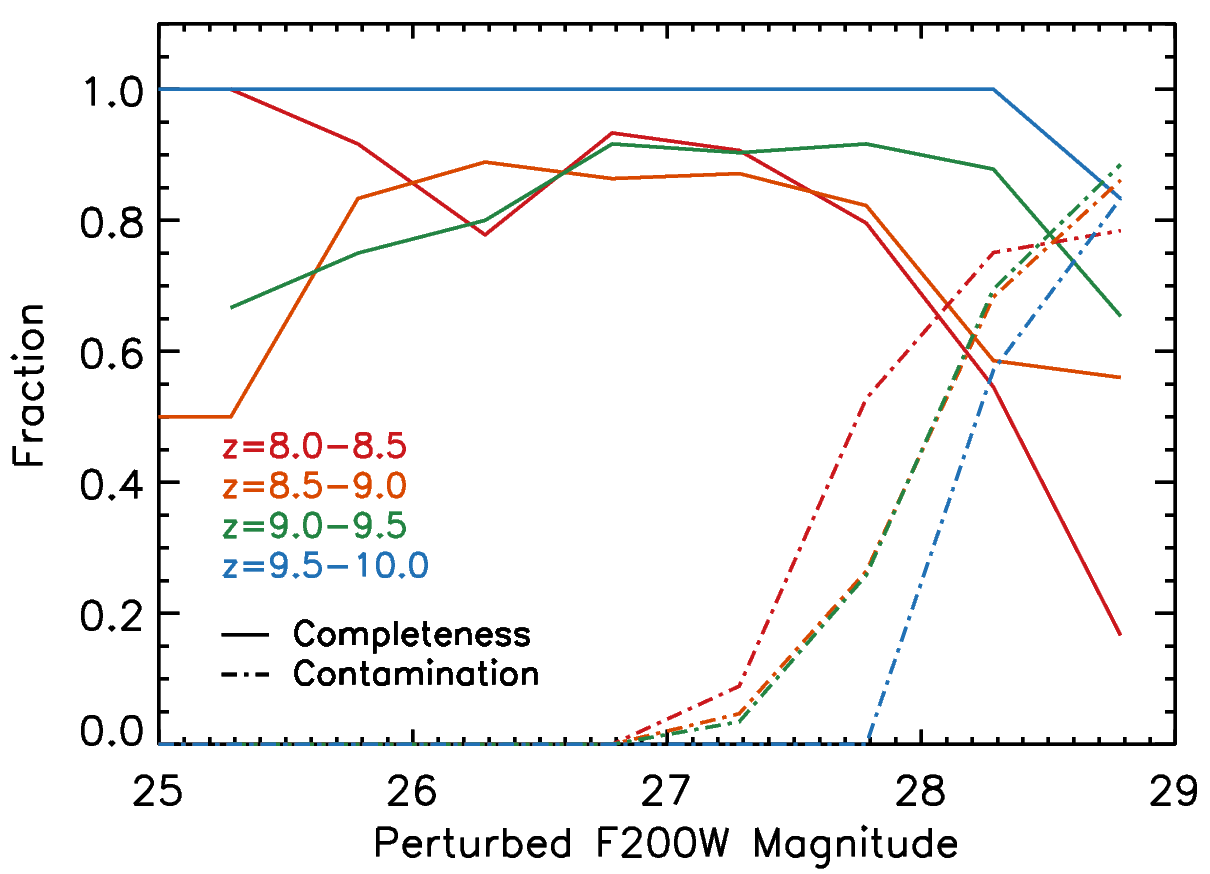}
    \caption{Here we plot our completeness and contamination fractions as a function of magnitude in several distinct redshift bins. As illustrated in Figure \ref{fig:finalsample}, we suffer from high contamination rates at the faint end of our sample (m$>$27.5) at all redshifts. It is also notable that the redshift range at which we recover the highest fraction of real $z>8$ galaxies is above $z=9.5$ where the \lya-break falls within the \jwst\ filters, providing the SED-fitting process the clearest high-redshift feature. Overall, we maintain a high recovery (completeness) fraction for our galaxies, where we recover a total of 897 of 1325 real $z>8$ sources in the SAM. The ones being missed by our selection criteria are predominantly at the faint end, close to our detection limits and where we are most dominated by contamination.}
    \label{fig:completecontam}
\end{figure}

Contaminants are defined as those galaxies that have an input redshift of $z<5$ but which meet all our selection criteria for a high-redshift ($z>8$) galaxy, and remain in our sample. The contamination fraction is the calculated as the total number of contaminants divided by the total number of sources in our final sample. In our final sample we have 2670 galaxies that met all of our selection criteria and 1294 of them are actual low-redshift contaminants giving a total contamination rate of $48.5\%$. In Figure \ref{fig:completecontam} we show our completeness fractions in different redshift bins as a function of magnitude (dashed lines) and note that just as shown in our final sample distribution (Figure \ref{fig:finalsample}) we become dominated by contamination at the fainter end, closer to our survey limit (at \textit{m} $>$ 28), but that contamination is very minimal at $m < 27.5$. We note also that these specific contamination rates are dependant on the colors of low-redshift galaxies in these simulations.  Finally, real galaxy surveys are sensitive to contamination from stellar sources, however the SAM only includes galaxies and as such we are exploring the impact of galactic contamination. 

It is important to note that these contamination rates are likely overestimates. They are dependent upon the colors of galaxies in the mock catalog at all redshifts.  Simulations in general struggle to produce redder galaxies at lower redshifts \citep[e.g.][]{somerville15, trayford16}; bluer overall colors for galaxies would lead to higher contamination rates in our sample. Additionally, \citet{yung19} described the need to reduce dust attenuation at higher-redshifts in the SAM we use for this paper. This primarily affects the simulated galaxies at $z > 4$, but our contaminating galaxies are typically at $z < 3$. Furthermore, the surface density of contaminants we measure for our sample is 1.65 arcmin$^{-2}$ which would imply $\sim$60 contaminating low-redshift interlopers in the 35 arcmin$^2$ of the first epoch of CEERS. This number is much greater than the total sample of candidate $z = 8.5-10$ galaxies observed in this field thus far \citep{donnan22, finkelstein22d}. Together, this implies our contamination rates are higher than we are likely to encounter in the real \jwst\ data.

\section{Conclusions}

Galaxies at $z > 8$ are expected to have bluer rest-frame UV colors than traditional model SED templates, which can lead to catastrophic photometric redshift failures. We explored the recommended FSPS templates included with the EAZY photometric-redshift fitting software \citep[][]{brammer08}, and found that they are all redder in the \jwst\ bands than the simulated $z>8$ galaxies from the CEERS mock catalogs \citet{yung22}. This is similar to what \citet{finkelstein22a} discovered for their observed $z=6-8$ galaxies. To enable improved photometric redshift measurements we created a supporting set of SED templates which match the predicted rest-UV colors of $z > 8$ simulated galaxies. We used EAZY to highlight the improvements in redshift recovery after the inclusion of our new template set, also suggested a set of criteria for selecting galaxies at $z > 8$ with \jwst\ surveys.
 
We use the published simulated galaxy catalog for CEERS as detailed in \citet{yung22}, which is based off of the Santa Cruz semi-analytic model (SAM) for galaxy formation \citep[][]{somerville99, somerville15} to which physically-predicted properties and star formation histories are assigned SEDs generated based upon SPS models from \citet{bruzual03}. This catalog contains a total of 1,472,791 galaxies between $z=0-10$, 6,578 of which are at $z>8$, but as real observations with \jwst\ are limited by our ability to detect objects in our data we impose a S/N $>3$ cut in F200W where the noise is set as the 1$\sigma$ depth of the CEERS observations. This leaves us with 913,288 simulated galaxies (3,084 at $z=8-10$) to use in determining the expected colors of high-redshift ($z>8$) galaxies as measured by \jwst. 
 
Our new suite of SED templates for fitting high-redshift ($z>8$) galaxies were designed to have properties expected of galaxies in the early Universe. We used the BPASS v2.2.1 \citep[][]{eldridge17, stanway18} model templates and selected for those that had low metallicity (5\% Z$_{\odot}$), young stellar populations (log stellar ages of 6, 6.5, and 7 Myr), inclusive of binary stars, and with an upper mass limit of 100 M$_{\odot}$ on a Chabrier IMF \citep[][]{chabrier03}. We note that these templates do not include any emission lines so we add another set of templates where we use {\sc Cloudy} v17.0 \citep[][]{ferland17} to model appropriate emission line spectra. In line with our expectations of high-redshift ($z>8$) galaxies we use high ionization parameters (log U = $-2$), low gas-phase metallicities ($Z=5\%Z_{\odot}$), Hydrogen gas density of 300 cm$^{-3}$ with a spherical covering fraction of 100\%, and remove \lya-emission as these galaxies are expected to be in a predominantly neutral IGM. These templates also include nebular continuum emission as well as emission lines, which produces redder colors for those templates with emission lines than those without. With this new set of six SED templates we are covering the full F200W $-$ F277W color space of our simulated high-redshift ($z>8$) galaxies (down to $-$0.43 mag), where the previous FSPS models only extended to $-$0.06 mag and where inclusion of the young, blue, low-mass galaxy, BX418 \citep[from][]{erb10} by previous studies such as \citet{finkelstein22a} had only reached a color of $-$0.2 mag. We make these templates publicly available for use at \href{https://ceers.github.io/LarsonSEDTemplates}{ceers.github.io/LarsonSEDTemplates}.
 
We also use our new suite of templates and the simulated CEERS catalog of galaxies to determine how best to select high-redshift ($z>8$) galaxies with \jwst, in ways that maximize completeness and minimize contamination by low-redshift ($z<5$) interlopers. What follows is the best criteria for identifying high-redshift candidates that we could determine for early \jwst\ data prior to having sufficient spectroscopic redshifts in this era to better calibrate our photometric redshift fits. We first make a requirement for a significant detection in both the F200W and F277W bands (S/N $>5$) to mimic detection bands in actual photometry. Then we require that the $\int$P( z > 7) > 0.85 from the EAZY redshift probability distribution, which removes any flat P(z)'s or those that have significant peaks at low redshift. We then place an upper limit on the $\chi^2$ of 15 to ensure a reasonably good fit to the data. We find that these criteria still leave our sample dominated by contamination by low-redshift ($z<5$) interlopers so we impose two color cuts: F150W $-$ F444W < 0.3 mag and F150W $-$ F200W $<$ 0.2 mag, dropping our overall contamination rate by $>$15\% while sacrificing only a handful of real high-redshift ($z>8$) sources. 

After applying these cuts, we find that our overall recovery rate of sources in our final sample that have input redshifts $z>8$ is over 65\%, only suffering from significant incompleteness at the faint end ($m>28$). This range is also where we encounter increased contamination fractions, though we expect the observed contamination rates to be lower than those predicted here. This is likely due to the simulated catalog not accurately reproducing the red colors of observed low-redshift galaxies.

We find that these above five selection criteria, combined with the inclusion of bluer SED templates such as the ones published here, are the best combination to ensure minimal contamination rates by low-redshift interlopers ($z<5$), while maximizing the recovery of real high-redshift ($z>8$). These results provide an important road map for observers venturing into this new era of astronomy with \jwst, while also highlighting the need for spectroscopic follow-up to confirm high-redshift galaxy candidates and measure accurate contamination rates.

\section*{Acknowledgements}
RLL, SLF, and MB acknowledge that they work at an institution, the University of Texas at Austin, that sits on indigenous land. The Tonkawa lived in central Texas and the Comanche and Apache moved through this area. We pay our respects to all the American Indian and Indigenous Peoples and communities who have been or have become a part of these lands and territories in Texas.  We are grateful to be able to live, work, collaborate, and learn on this piece of Turtle Island.  

This material is based upon work supported by the National Science Foundation Graduate Research Fellowship under Grant No. DGE-1610403.  RLL and SLF acknowledge support from NASA through ADAP award 80NSSC18K0954. AY and TAH are supported by an appointment to the NASA Postdoctoral Program (NPP) at NASA Goddard Space Flight Center, administered by Oak Ridge Associated Universities under contract with NASA.

The authors acknowledge the Texas Advanced Computing Center (TACC) at The University of Texas at Austin for providing database, and grid resources that have contributed to the research results reported within this paper. \href{http://www.tacc.utexas.edu}{www.tacc.utexas.edu}


\vspace{5mm}
\facilities{Texas Advanced Computing Center}

\software{BPASS \citep[][]{eldridge17,stanway18},  
          {\sc Cloudy} v17.0 \citep{ferland17}, 
          EAZY \citep{brammer08},
          IDL Astronomy Library: \url{idlastro.gsfc.nasa.gov} \citep{Landsman93}}

\bibliography{main}{}
\bibliographystyle{aasjournal}

\end{document}